\pdfoutput=1
\documentclass[11pt,twoside,a4paper,cmspaper,final,collab]{cms-tdr}
\def\svnVersion{5cef060}\def\svnDate{2026/11/30}\def\cmsCernNoTag{CERN-EP-2026-157}\def\cmsCernDate{\today}\def\cmsMessage{Published in Physical Review Letters as \href{http://dx.doi.org/10.1103/2zxk-9yvw}{\doi{10.1103/2zxk-9yvw}.}}
\begin{document}\cmsNoteHeader{HIG-24-012}

\newlength\cmsTabSkip\setlength{\cmsTabSkip}{0.3ex}
\newlength\cmsBigSkip\setlength{\cmsBigSkip}{0.6ex}

\newlength\cmsFigWidth

\ifthenelse{\boolean{cms@external}}{%
    \providecommand{\cmsLeft}{upper\xspace}
    \providecommand{\cmsRight}{lower\xspace}
    \setlength{\cmsFigWidth}{\columnwidth}
    \def\appendixname{End Matter}
    \providecommand{\EndMatter}{ in the \appendixname\xspace}

}{
    \providecommand{\cmsLeft}{left\xspace}
    \providecommand{\cmsRight}{right\xspace}
    \setlength{\cmsFigWidth}{0.7\columnwidth}
    \providecommand{\EndMatter}{ in the \nameref{endmatter}\xspace}

}

\ifthenelse{\boolean{cms@external}}{
    \providecommand{\cmsTable}[1]{\resizebox{\columnwidth}{!}{#1}}
}
{
    \providecommand\cmsTable[1]{#1}
}

\newcommand{\pp}{\ensuremath{\Pp\Pp}\xspace}
\newcommand{\prC}{\ensuremath{{\mathrm{C}}}}
\newcommand{\prB}{\ensuremath{{\mathrm{B}}}}
\newcommand{\HHbbbb}{\ensuremath{\PH\PH\to 4\PQb}\xspace}
\newcommand{\HH}{\ensuremath{\PH\PH}\xspace}
\newcommand{\PNet}{\textsc{ParticleNet}\xspace}
\newcommand{\SPANET}{\textsc{SPANet}\xspace}
\newcommand{\DeepJet}{\textsc{DeepJet}\xspace}
\newcommand{\DeepCSV}{\textsc{DeepCSV}\xspace}
\newcommand{\Prob}{\ensuremath{\mathrm{Prob}}\xspace}
\newcommand{\HHH}{\ensuremath{{\PH\PH\PH}}\xspace}
\newcommand{\ttHH}{\ensuremath{{\ttbar\PH\PH}}\xspace}
\newcommand{\ttH}{\ensuremath{{\ttbar\PH}}\xspace}
\newcommand{\VHH}{\ensuremath{{\PV\PH\PH}}\xspace}
\newcommand{\VH}{\ensuremath{{\PV\PH}}\xspace}
\newcommand{\HHHdecay}{\ensuremath{\PH\PH\PH\to 6\PQb}\xspace}
\newcommand{\Probttincl}{\ensuremath{\Prob_{\ttbar}^\text{incl}}\xspace}
\newcommand{\ProbHHHincl}{\ensuremath{\Prob_{\HHH}^\text{incl}}\xspace}
\newcommand{\ProbHH}{\ensuremath{\Prob_{\HH}}\xspace}
\newcommand{\ProbHHH}{\ensuremath{\Prob_{\HHH}}\xspace}
\newcommand{\ProbQCD}{\ensuremath{\Prob_{\mathrm{QCD}}}\xspace}
\newcommand{\Probtt}{\ensuremath{\Prob_{\ttbar}}\xspace}
\newcommand{\ProbMultiH}{\ensuremath{\mathrm{ProbMultiH}}\xspace}
\newcommand{\klambda}{\ensuremath{\kappa_3}\xspace}

\title{
    Search for nonresonant triple Higgs boson production in the final state with six bottom quarks in proton-proton collisions at \texorpdfstring{$\sqrt{s}=13\TeV$}{sqrt(s) = 13 TeV}
}

\cmsNoteHeader{HIG-24-012}

\author*[cern]{The CMS Collaboration}

\date{\today}

\abstract{
    A search for nonresonant triple Higgs boson (\HHH) production in the final state with six bottom quarks is performed using proton-proton collisions at $\sqrt{s}=13\TeV$ corresponding to an integrated luminosity of 138\fbinv recorded by the CMS experiment. No significant excess of events over the background prediction is seen. Observed (expected) 95\% confidence level upper limits on the signal cross section are set at 44 (43)\unit{fb}, corresponding to 588 (572) times the standard model expectation. The \HHH process is directly sensitive to both the trilinear and quartic Higgs self-couplings. The observed (expected) constraint on the quartic coupling modifier $\kappa_4$ is $-177 < \kappa_4 < 185$ ($-180 < \kappa_4 < 190$), assuming $\kappa_3=1$. The corresponding constraint on the trilinear coupling modifier $\kappa_3$ is $-7.4 < \kappa_3 < 12.4$ ($-6.4 < \kappa_3 < 11.2$), assuming $\kappa_4=1$. This analysis provides the most stringent constraint to date on nonresonant \HHH production and excludes, for the first time, part of the $(\kappa_3,\kappa_4)$ space allowed by the perturbative unitarity bound.
}

\hypersetup{
    pdfauthor={CMS Collaboration},
    pdftitle={Search for the nonresonant triple Higgs boson production in the final state with six bottom quarks in proton-proton collisions at sqrt(s) = 13 TeV},
    pdfsubject={CMS},
    pdfkeywords={CMS, HHH, 6b},
}

\maketitle

Following the discovery of the Higgs boson (\PH)~\cite{PhysRevLett.13.321,Higgs:1964ia,PhysRevLett.13.508,PhysRevLett.13.585,PhysRev.145.1156,PhysRev.155.1554,ATLAS:2012yve,CMS:2012qbp,CMS:2013bcl}, probing the structure of the Higgs potential has become a central goal of the LHC physics program.
In the standard model (SM), the potential is determined by the vacuum expectation value $v \approx 246\GeV$ and the \PH mass $m_\PH = 125\GeV$, leading to trilinear ($\lambda_3$) and quartic ($\lambda_4$) self-couplings, both equal to $\lambda_\text{SM} = m_\PH^2 / (2v^2)$.
Deviations from this prediction could have profound consequences for the early universe and its stability~\cite{Horn:2020wif,PhysRevD.97.075008,PhysRevD.78.063518,Markkanen:2018pdo}.

At the LHC, $\lambda_3$ and $\lambda_4$ can be directly probed via \PH pair (\HH) and triple \PH (\HHH) production.
In proton-proton ($\Pp\Pp$) collisions at $\sqrt{s}=13\TeV$, the dominant production mechanism for both processes is gluon-gluon fusion (ggF), with cross sections of $\sigma^{\HH}_\text{ggF}=31^{+2}_{-7}$\unit{fb}~\cite{Grazzini:2018bsd,PhysRevD.58.115012,PhysRevLett.117.012001,Baglio:2018lrj,PhysRevLett.111.201801,Shao:2013bz,deFlorian:2015moa,PhysRevD.103.056002,Heinrich:2022idm,Bagnaschi:2023rbx} and $\sigma^{\HHH}_\text{ggF} = 0.08\pm0.01\unit{fb}$~\cite{Maltoni:2014eza,Papaefstathiou:2023uum,deFlorian:2019app,Abouabid:2024gms}, respectively, calculated at next-to-next-to-leading order (NNLO).
Deviations from the SM predictions are parameterized by the \PH self-coupling modifiers, defined as $\kappa_3=\lambda_3/\lambda_\text{SM}$ and $\kappa_4=\lambda_4/\lambda_\text{SM}$, respectively, following the convention used in the $\kappa$ framework~\cite{LHCHiggsCrossSectionWorkingGroup:2013rie}.
Both the \HH and \HHH processes are under active investigation at the LHC, with searches in multiple final states~\cite{ATLAS:2023gzn,PhysRevD.108.052003,PhysRevD.110.032012,ATLAS:2024lhu,ATLAS:2023elc,CMS:2022hgz,CMS:2024fkb,CMS:2022kdx,CMS:2024rgy,PhysRevLett.133.101801,CMS:2022dwd,ATLAS:2024lsk,PhysRevLett.129.081802,CMS:2022gjd,PhysRevD.111.032006}.
Although \HH production is primarily sensitive to $\kappa_3$, it also depends on $\kappa_4$ through electroweak corrections. This dependence is subleading for LHC energies~\cite{Bizon:2018syu,Bizon:2024juq}, and no dedicated extraction of $\kappa_4$ from \HH measurements is currently available.
The ATLAS and CMS experiments have set 95\% confidence level (\CL) upper limits on the \HH signal-strength modifier, defined as $\mu_{\HH}=\sigma_{\HH}/\sigma_{\HH}^{\text{SM}}$, of $\mu_{\HH}<2.9$ and $<3.5$, respectively, assuming SM production~\cite{PhysRevLett.133.101801,CMS:2022dwd,CMS-PAS-HIG-20-011}.
The $\kappa_3$ coupling modifier is constrained by ATLAS and CMS to be $-1.2 < \kappa_3 < 7.2$~\cite{PhysRevLett.133.101801} and $-1.4 < \kappa_3 < 7.0$~\cite{CMS:2022dwd,CMS-PAS-HIG-20-011} at 95\% \CL, respectively.
The \HHH production provides direct sensitivity to both $\kappa_3$ and $\kappa_4$, but suffers from a small cross section due to destructive interference among contributing diagrams.
The ATLAS Collaboration reported an observed (expected) limit on the signal strength, relative to the SM, of $\mu_{\HHH}=760$ (750) in the $\PH\PH\PH \to 6\PQb$ final state using 139\fbinv of data at $\sqrt{s}=13\TeV$~\cite{PhysRevD.111.032006}, with coupling modifier constraints $-11 < \kappa_3 < 17$ for $\kappa_4 = 1$ and $-230 < \kappa_4 < 240$ for $\kappa_3 = 1$.
Perturbative unitarity of the $\HH\to\HH$ scattering imposes bounds on $\kappa_3$ and $\kappa_4$, thereby delineating the region where interpretations remain theoretically consistent~\cite{Stylianou:2024epjc}.
Extended Higgs sectors featuring a strong first-order electroweak phase transition, potentially enabling electroweak baryogenesis, can produce large deviations of the \HHH cross section with respect to the SM prediction, with modest modifications to the \HH cross section, while ``spontaneous flavor violation'' provides additional well-motivated scenarios with distinctive \PH phenomenology~\cite{Biermann:2024oyy,Egana-Ugrinovic:2018znw,Egana-Ugrinovic:2019dqu}.
The \HHH production has also been studied as a probe of the \PH self-couplings at future 100\TeV $\Pp\Pp$ colliders~\cite{Fuks:2015hna,Papaefstathiou:2019ofh}.

This Letter presents a search for nonresonant $\PH\PH\PH \to 6\PQb$ production, using $\Pp\Pp$ collision data collected by the CMS experiment between 2016 and 2018 at $\sqrt{s} = 13\TeV$, corresponding to an integrated luminosity of 138\fbinv~\cite{CMS:2021iwu,CMS:2018elu,CMS:2019lum}.
The CMS apparatus~\cite{CMS:2008xjf,CMS:2023gfb} is a multipurpose, nearly hermetic detector, designed to trigger on~\cite{CMS:2020cmk,CMS:2016ngn,CMS:2024aqx} and identify electrons, muons, photons, and charged and neutral hadrons~\cite{CMS:2020uim,CMS:2018rym,CMS:2014pgm}.
The ggF \HHH signal is generated at leading order (LO) using \MGvATNLO~\cite{Alwall:2014hca} and normalized to the NNLO cross section; the irreducible ggF \HH background is simulated at next-to-LO (NLO) with the \POWHEG program~\cite{POWHEG1,POWHEG2,POWHEG3,Bagnaschi:2011tu,Heinrich:2017kxx,Jones:2017giv,Buchalla:2018yce,Heinrich:2019bkc,Heinrich:2020ckp,Davies:2019dfy,Bagnaschi:2023rbx} and normalized to NNLO.
Parton showering and hadronization are modeled with \PYTHIA8.205~\cite{Sjostrand:2014zea} using the CP5 tune~\cite{CMS:2019csb} and NNPDF3.1 NNLO parton distribution functions (PDFs)~\cite{Ball:2017}.
Events are reconstructed with the particle-flow algorithm~\cite{CMS:2017yfk}.
The effects of additional $\Pp\Pp$ interactions in the same or neighboring bunch crossings (pileup) are mitigated using charged-hadron subtraction for small-radius jets~\cite{CMS-PAS-JME-14-001} and the PUPPI algorithm for large-radius jets~\cite{Bertolini:2014bba,Sirunyan:2020foa}.
Jets are clustered with the anti-\kt algorithm~\cite{Cacciari:2008gp} using distance parameters 0.4 (small-radius) and 0.8 (large-radius), as implemented in \FASTJET~\cite{Cacciari:2011ma}, and \PQb jets are identified with the \PNet tagger~\cite{Qu:2019gqs,CMS-DP-2025-009}, with performance calibrated as in Ref.~\cite{CMS:2025tcl}.
Large-radius \PH candidates from $\PH\to\bbbar$ decays are tagged with the \PNet \bbbar discriminator, calibrated as in Ref.~\cite{CMS:2025xkn}.
Events are selected with multijet triggers~\cite{CMS:2020cmk,CMS:2016ngn,CMS:2024aqx}, giving an acceptance of $\sim$80--90\% for signal events in the analysis phase space.
The offline selection requires at least four small-radius jets with $\pt > 25\GeV$ and $\abs{\eta} < 2.4$--2.5, depending on the data-taking period, and large-radius jets with $\pt > 300\GeV$ and $\abs{\eta} < 2.5$.
Further details of the signal and background simulated samples, detector subsystems, trigger requirements, and object definitions are provided\EndMatter.
Tabulated results are provided in the HEPData record for this analysis~\cite{hepdata}.

Owing to the large branching fraction of $\PH\to\bbbar$ ($\sim$60\%), this final state offers a rich topology in which each \PH is reconstructed either from two small-radius jets (resolved) or a single large-radius jet (merged).
To maximize sensitivity, novel \PH tagging techniques, as well as an attention-based deep neural network with a transformer architecture that enables inclusive identification of signal events across all combinations of resolved and merged \PH candidates, are employed.
An inclusive strategy is employed and targets four distinct reconstruction topologies of the \PH candidates: \emph{resolved}, where all three \PH are reconstructed from pairs of small-radius jets; \emph{partially resolved} and \emph{partially merged}, where, respectively, one or two \PH are reconstructed from a single large-radius jet and the remainder from small-radius jets; and \emph{merged}, where all three \PH are reconstructed from large-radius jets.

A symmetry-preserving attention network (\SPANET)~\cite{Shmakov:2021qdz,Fenton:2022PhysRevD,Fenton:2024CommPhys} is trained on simulated samples to separate the signal from the main backgrounds: quantum chromodynamics (QCD) multijet, top quark-antiquark (\ttbar), and \HHbbbb production.
The same network is also used to categorize the signal among distinct topologies by assigning or combining jets to form \PH candidates.
The \SPANET algorithm addresses the jet assignment problem by explicitly preserving the permutation symmetry of identical final-state particles, and it has recently been shown to be potentially effective for both \HH and \HHH event reconstruction~\cite{Chiang:2024pho,Li:2024qfq}.
It yields per-jet assignment probabilities and topology scores that are used to define signal-enriched categories.
The features input to \SPANET include the transverse momentum \pt, pseudorapidity $\eta$, $\cos\phi$, $\sin\phi$ (where $\phi$ is the azimuthal angle), and mass of up to ten small-radius and three large-radius jets, ordered by decreasing \PQb- and \bbbar-tagging scores.
The corresponding discretized \PNet \PQb and \bbbar tagging scores are also included.
Small- and large-radius jets are both retained as network inputs, including spatially overlapping jets.
For training-target definition and object calibration, overlaps are resolved by prioritizing large-radius jets, which avoids double counting and reduces combinatorics.
Additional inputs comprise jet pairwise angular separations, dijet masses, and event-level variables, such as the scalar jet \pt sum \HT and missing transverse momentum \ptmiss.
The data-taking period is also included as a discrete input feature to account for differences in conditions, including trigger configurations, detector performance, and pileup.

The classification proceeds in two stages to enhance the sensitivity to the $\PH\PH\PH \to 6\PQb$ signal.
The first stage, referred to as the inclusive classifier, is trained on simulated events passing the trigger selection and containing at least four small-radius jets to classify them into four categories: $\PH\PH\PH \to 6\PQb$, $\PH\PH \to 4\PQb$, QCD multijet, and $\ttbar$.
Its normalized outputs, $\Prob_{j}^{\text{incl}}$, can be interpreted as the probabilities for an event to originate from process $j$.
Events for which $\ProbHHHincl$ is the largest of the four outputs are selected, retaining $\sim$80\% of $\PH\PH\PH \to 6\PQb$ signal events.

The selected events are then passed to a second classifier, trained on this signal-enriched subset using the same architecture, hyperparameters, and target classes as the first stage to further refine the identification of the target topology.
In this selected sample, the background is dominated by QCD multijet production ($\sim$95\%), with \ttbar contributing the remaining $\sim$5\%.
The outputs are interpreted as class probabilities \ProbHHH, \ProbHH, \ProbQCD, and \Probtt (normalized to sum to unity).
Events for which \ProbHHH is the largest of the four second-stage outputs are retained.
Relative to the first-stage-selected sample, this requirement retains $\sim$90\% of the $\PH\PH\PH \to 6\PQb$ signal while reducing the QCD multijet, \ttbar, and $\PH\PH \to 4\PQb$ backgrounds by additional factors of 4, 2, and 1.3, respectively.
This two-stage approach is effectively a boosting ensemble method~\cite{Schapire2003}: the first network performs a coarse event selection, and the second focuses on the signal-enriched sample, adapting to the remaining difficult-to-classify events and thereby improving the overall discrimination power.
Using the second classifier alone would require retraining it on the full event sample, since its training phase space is defined by the first-stage selection.
No further gain is observed from additional classifier stages.

In simulation, an \PH is considered reconstructible if its decay products are matched to either two selected small-radius jets or one selected large-radius jet within the tracker acceptance.
With this definition, 27\% of signal events have three reconstructible \PH (3\PH), 52\% have two (2\PH), and 21\% have one (1\PH) or zero (0\PH).
Considering the classifications according to reconstructible \PH, a third deep neural network, trained with optimized hyperparameters using the same input features, assigns each event to one of ten categories, defined by the number of merged \PH (mh) and resolved \PH (h): 0mh0h, 0mh1h, 1mh0h, 2mh0h, 1mh1h, 0mh2h, 3mh0h, 2mh1h, 1mh2h, and 0mh3h.
The seven 3\PH and 2\PH categories form the signal regions (SRs), while the three 1\PH and 0\PH categories define control regions (CRs) used to assess the agreement between data and the background model.

Signal extraction is performed via a binned maximum likelihood fit to the $\ProbMultiH \equiv \ProbHH + \ProbHHH$ distribution.
In each category, only the region with $\ProbMultiH>0.8$ is considered and divided into ten bins.
Together, the two classifier selections, assignment to an SR category, and the $\ProbMultiH>0.8$ requirement retain 56\% of the $\PH\PH\PH \to 6\PQb$ signal events entering the first-stage classification.
The most sensitive bin is defined such that the relative statistical uncertainty originating from the background model is $\sim$20\%.
The remaining nine bins are evenly spaced in \ProbMultiH.

At high values of \ProbHHH, the kinematic properties of the QCD multijet and \ttbar backgrounds mimic those of the \HHH and \HH signals.
Many QCD multijet events exhibit signal-like kinematics, but fail the selection because of their jet flavor composition.
To model the background in the SRs, data events failing the inclusive \HHHdecay and \HHbbbb classifications are used.
For each event in the background model, the complete set of discretized \PQb and \bbbar tagging scores is sampled jointly from the distribution observed in events for which \ProbHHH is the largest of the four classifier outputs and transferred to the small- and large-radius jets, preserving the multiplicity patterns and correlations of high tagging scores within and between the two jet collections.
In each category, the normalization of this background model is fixed to the corresponding data yield.
The \SPANET inference is applied to these embedded events, yielding an inclusive description of all backgrounds.
The method is validated in categories enriched with $\PH\PH \to 4\PQb$ events, where $\PH\PH\PH \to 6\PQb$ contamination is negligible, and in sidebands of the multijet mass with respect to the 3\PH expectation.
It is found to reproduce the data within uncertainties.
The relative statistical uncertainty in the background model is bin- and category-dependent, ranging from below 1\% in the most populated sideband bins to approximately 20\% in the most sensitive bin.
A separate \ttbar CR is included, defined by events with high predicted \ttbar score \Probttincl and a purity of $\sim$99\%.
Agreement between data and \ttbar simulation validates the analysis strategy and the jet-tagging calibrations.

While the ggF \HH contribution is nonnegligible in the SR, particularly for constraints on \PH self-couplings, other \PH and \HH production modes are negligible even under variations of the trilinear coupling modifier \klambda.
These include \HH production via vector boson fusion, and \PH and \HH production in association with top quarks (\ttH, \ttHH) or electroweak vector bosons (\VH, \VHH).
Their yields depend only weakly on $\kappa_3$ and $\kappa_4$~\cite{new9,Maltoni:2017ims,Haisch:2025pql}, and they are therefore omitted from the signal extraction.

Three primary sources of systematic uncertainties are considered: theoretical uncertainties, uncertainties related to corrections affecting the \HHH and \HH signal templates, and uncertainties in the background estimate using control samples in data.
Theoretical uncertainties in the \HHH and \HH signals are evaluated through renormalization and factorization scales ($\sim$20\%), PDF ($\sim$3\%), and branching fraction ($\sim$1\%) variations.
These affect the overall normalization and induce small shape variations through bin migration driven by changes in jet kinematics.

Jet-related uncertainties, including the energy scale, energy resolution, and mass resolution, affect jet and dijet kinematics and hence the \SPANET prediction.
For each variation, the event classification and categorization score is recalculated, and new signal templates are constructed to assess both shape and normalization uncertainties.
Changes in event selection efficiency due to these variations lead to normalization shifts.
Corrections related to flavor tagging result in a $\sim$15\% reduction of the \HHH signal yield in the highest \ProbMultiH bins.
Associated uncertainties are propagated to the signal templates following the prescription of Ref.~\cite{CMS:2025tcl}.
For small-radius jets, uncertainties are decomposed into sources corresponding to data-to-simulation scale factors (SFs) for \PQb tagging derived from the $\PZ$+jets, $\PW$+jets, and \ttbar processes, as well as modeling effects from parton showering, \PQb fragmentation, pileup, integrated luminosity, and jet-related uncertainties.
For large-radius jets, a single \bbbar tagging uncertainty, corresponding to that in the data-to-simulation SF measured in $\PZ\to\bbbar$ events, is applied.
All sources are treated as uncorrelated and implemented as event weight variations.
The dominant uncertainty arises from the statistical precision of the \PQb-tagging calibration on small-radius jets, leading to a $\sim$5\% uncertainty in the yield.
For large-radius jets, the \bbbar tagging uncertainty in the yield is $\sim$22\%.

The uncertainty in the QCD multijet background estimate based on control samples in data accounts for potential mismodeling beyond statistical fluctuations.
As an extreme variation, the shape of the background model in the 1\PH category is assigned as an alternative background model in the \HHH SR and treated as uncorrelated across categories.
This decorrelated shape variation allows the fit to adjust for mismodeling in the lower-score bins of the falling \ProbMultiH distribution.

The precision of the analysis is primarily limited by data statistical uncertainties, which account for $\sim$52\% of the nominal expected limit.
The dominant systematic uncertainty arises from the QCD multijet background model, including its shape uncertainty and the per-bin statistical variance.
Statistical uncertainties in the simulated \HHH and \HH templates are also included in the fit~\cite{Barlow:1993dm}.
Theory uncertainties in the signal and experimental uncertainties from jet reconstruction and flavor identification are subdominant.

The contribution from the \HH process in the SR is suppressed by requiring \ProbHHH to be the largest of the four classifier outputs.
In the SR, the expected \HH yield is about one order of magnitude larger than the \HHH yield for the SM hypothesis; this hierarchy is even stronger for anomalous couplings because the \HH cross section increases more rapidly than the \HHH cross section~\cite{Bizon:2024juq}.
Both \HHH and \HH production are included in the statistical model.
For the \HHH signal-strength interpretation, the \HH contribution is fixed to its SM prediction and its systematic uncertainties are profiled.
For the coupling interpretation, the \HHH and \HH normalizations are varied as functions of $\kappa_3$ and $\kappa_4$.
The SM top quark Yukawa coupling is assumed.

\begin{figure}[ht]
    \centering
    \includegraphics[width=\cmsFigWidth]{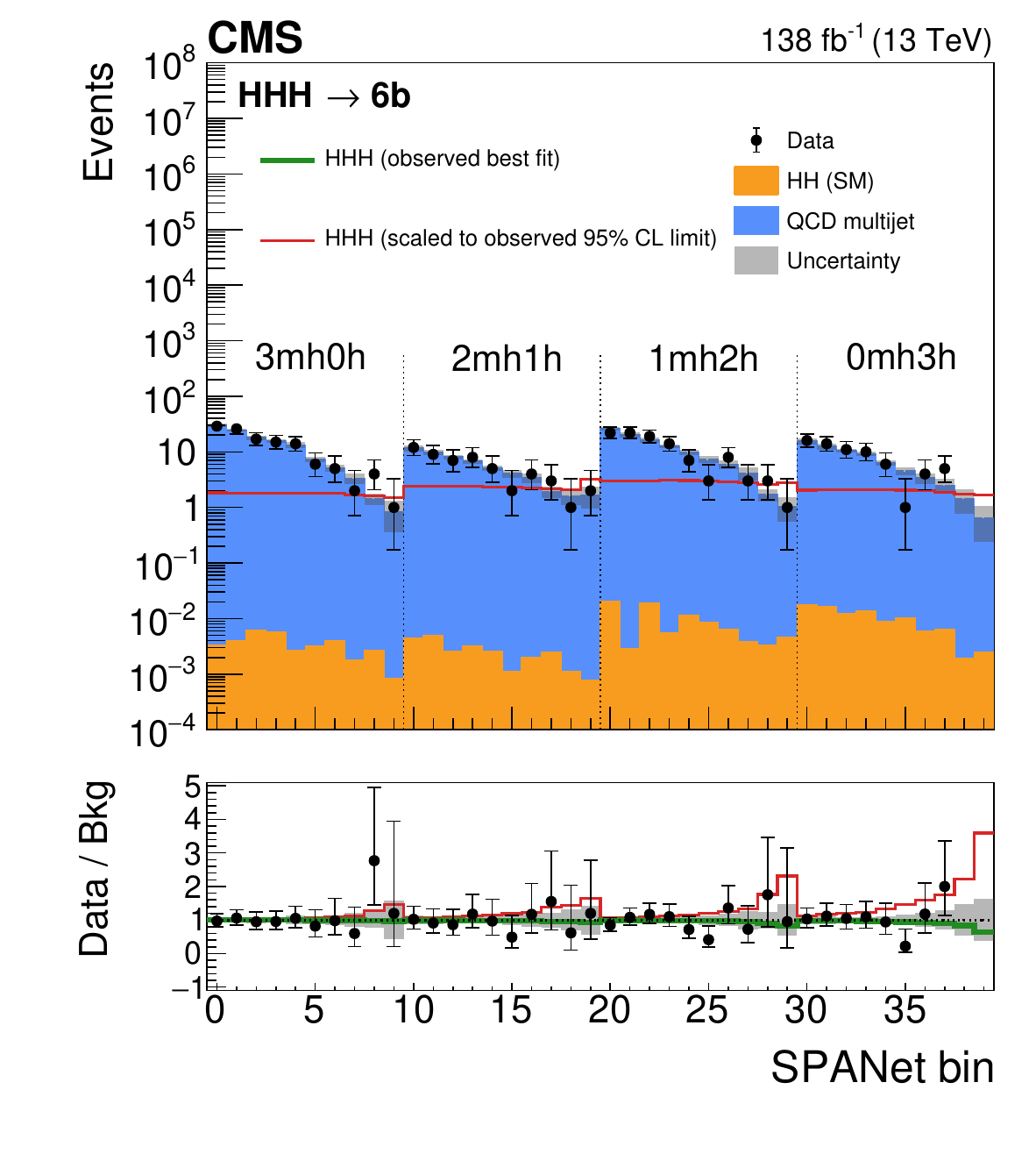}
    \caption{
        Postfit yields in the SR categories with three reconstructed \PH.
        The \HH contribution is fixed to its SM prediction; the green and red curves show, respectively, the observed best-fit \HHH contribution and the \HHH signal normalized to the observed 95\% \CL UL of 588 times the SM prediction.
        Bins correspond to ten intervals per category of the \SPANET score \ProbMultiH in the high-score range 0.8--1.0, ordered by increasing purity.
    }
    \label{fig:postfit_Bonly-run2_HH_separate}
\end{figure}

No evidence for an \HHH signal is observed, and two interpretations are presented: an upper limit (UL) on the SM-like \HHH production rate and constraints in the $(\kappa_3, \kappa_4)$ parameter space.
The postfit distributions in the 3\PH SR are shown in Fig.~\ref{fig:postfit_Bonly-run2_HH_separate}.

\begin{figure}[htbp]
    \centering
    \includegraphics[width=0.49\textwidth]{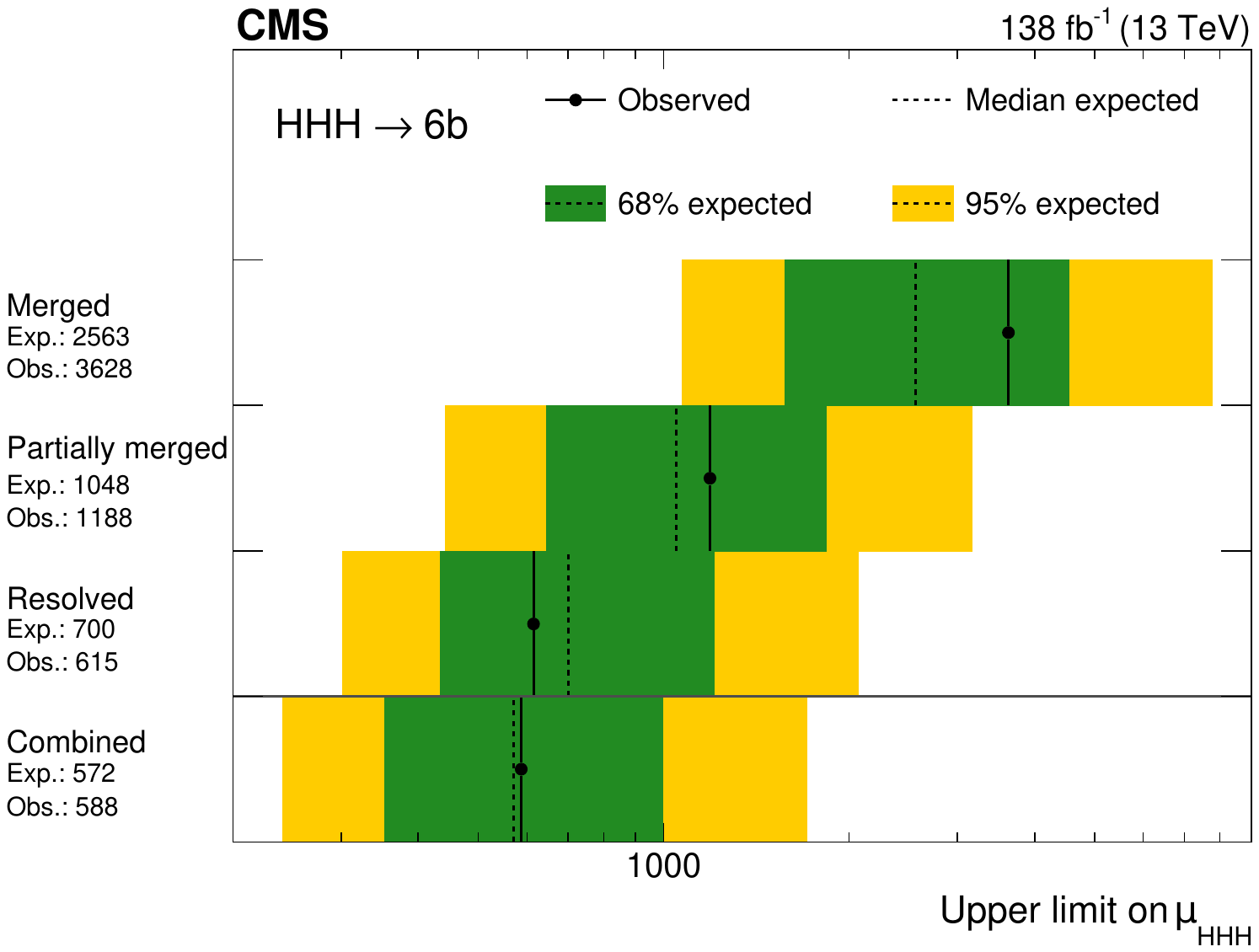}
    \includegraphics[width=0.49\textwidth]{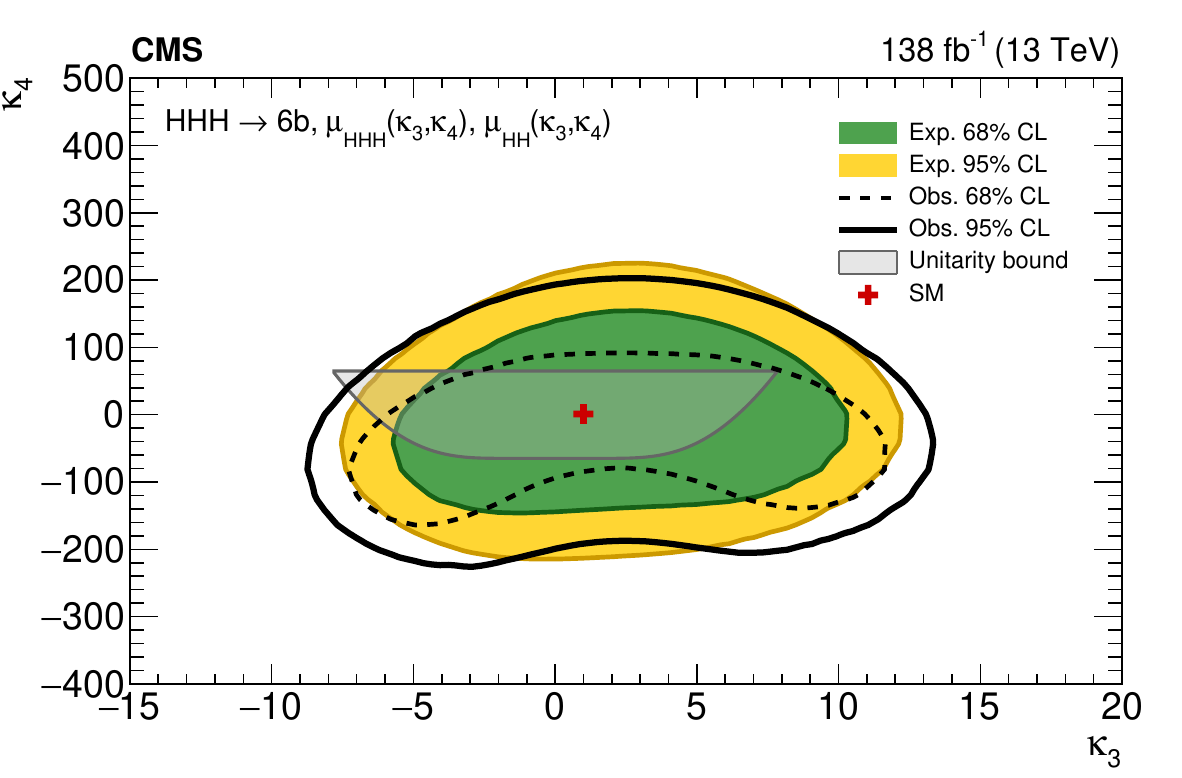}
    \caption{
        Expected and observed ULs on the \HHH signal strength shown separately for resolved, partially resolved and partially merged, and merged event topologies, along with the combined result (\cmsLeft).
        The 68 and 95\% \CL contours in the $(\kappa_3,\kappa_4)$ plane from the combined profile likelihood scan (\cmsRight).
        The gray region denotes parameter space allowed by perturbative unitarity in $\HH\to\HH$ scattering~\cite{Stylianou:2024epjc}.
    }
    \label{fig:results-summary}
\end{figure}

The ULs at 95\% \CL on the \HHH signal strength are derived based on the asymptotic formulae for the profile likelihood ratio test statistic $-2\Delta\ln(L)$~\cite{LHCCLs, Cowan:2010js} and the \CLs~\cite{Junk:1999kv,Read:2002hq} criterion.
The results have been obtained using the CMS statistical analysis tool \textsc{Combine}~\cite{CMS:2024onh}, which is based on the \textsc{RooFit}~\cite{Verkerke:2003ir} and \textsc{RooStats}~\cite{Moneta:2010pm} frameworks.
Figure~\ref{fig:results-summary} (\cmsLeft) shows the resulting limits across resolved, partially resolved and partially merged, and merged event topologies.
The resolved categories provide the highest sensitivity.
Partially resolved, partially merged, and merged topologies enhance the overall sensitivity by up to 20\% relative to the resolved categories alone.
While the 3\PH region alone yields the strongest constraint among individual groups, the combination of all categories improves the sensitivity by $\sim$10\%.
The observed (expected) UL on the \HHH production signal strength is 588 (572) times the SM prediction at 95\% \CL, corresponding to a UL on the cross section of 44 (43)\unit{fb}.

To constrain the \PH self-coupling modifiers, two types of profile likelihood scans are performed: one-dimensional (1D) scans, in which one coupling modifier is varied while the other is fixed to its SM value, and 2D scans, in which both couplings are varied simultaneously.
The 1D scans provide the most stringent constraints on $\kappa_4$ and $\kappa_3$.

The observed (expected) constraints are $-177 < \kappa_4 < 185$ ($-180 < \kappa_4 < 190$) assuming $\kappa_3 = 1$, and $-7.4 < \kappa_3 < 12.4$ ($-6.4 < \kappa_3 < 11.2$) assuming $\kappa_4 = 1$, at 95\% \CL.
The $\kappa_3$ sensitivity is primarily driven by the dependence of \HHH production on $\kappa_3$.
Including the $\kappa_3$-dependent \HH contribution improves the expected constraint by $\sim$9\%.
Two-parameter likelihood scans in the $(\kappa_3,\kappa_4)$ plane yield the 68 and 95\% \CL contours shown in Fig.~\ref{fig:results-summary} (\cmsRight).
For \HHH and \HH production, both $\kappa_3$ and $\kappa_4$ are varied following the parametrization in Refs.~\cite{Bizon:2018syu,Ding:2951745}.
Notably, the simultaneous $(\kappa_3,\kappa_4)$ fit yields expected and observed constraints that exclude regions of parameter space permitted by the perturbative unitarity bound~\cite{Stylianou:2024epjc}.
Complementary projected sensitivities for the high-luminosity (HL)-LHC scenario ($\sqrt{s}=14\TeV$, 3\abinv) are presented\EndMatter.
The coupling scans vary the \HHH and \HH cross sections as functions of $\kappa_3$ and $\kappa_4$ while using signal templates and selection efficiencies evaluated at $\kappa_3 = \kappa_4 = 1$.
Coupling-dependent kinematic effects are therefore not included in the nominal scans.
An event-reweighting study following Refs.~\cite{PhysRevLett.129.081802,CMS-PAS-HIG-20-011} estimates that these effects degrade the expected $\kappa_3$ boundary by 12--15\% for $\kappa_4 = 1$.
A fully simulated \HHH grid is used to assess the kinematic effects across the two-dimensional plane.
In the region where the contour probes the perturbative unitarity bound, the expected boundary changes by approximately 2--4\%, leaving the exclusion stable.
Simultaneous $\kappa_3$- and $\kappa_4$-dependent kinematic effects in \HH production have been studied in recent references, including through an event-generation implementation~\cite{new5,new6,Abouabid:2024gms}.
Their incorporation requires dedicated validation in the phase space of both \HH and \HHH analyses and is beyond the scope of this analysis.

In summary, this Letter presents a search for nonresonant triple Higgs boson (\HHH) production in the final state with six bottom quarks using 138\fbinv of proton-proton collision data at $\sqrt{s}=13\TeV$ collected by the CMS experiment between 2016 and 2018.
No significant deviation from the background prediction is observed.
At 95\% confidence level, the observed (expected) upper limit on the \HHH production cross section is 44 (43)\unit{fb}, corresponding to 588 (572) times the standard model prediction.
In the framework of Higgs boson self-coupling modifiers ($\kappa$), the search constrains the quartic coupling modifier $\kappa_4$ to $-177 < \kappa_4 < 185$ for $\kappa_3=1$ and constrains the trilinear coupling modifier $\kappa_3$ to $-7.4 < \kappa_3 < 12.4$ for $\kappa_4=1$.
The two-dimensional $(\kappa_3,\kappa_4)$ scan constrains, for the first time in a direct search for nonresonant \HHH production, the perturbative unitarity boundary and excludes part of the allowed parameter space.
These results are the most stringent constraints to date on nonresonant \HHH production and on $\kappa_4$.

\begin{acknowledgments}
    We congratulate our colleagues in the CERN accelerator departments for the excellent performance of the LHC and thank the technical and administrative staffs at CERN and at other CMS institutes for their contributions to the success of the CMS effort. In addition, we gratefully acknowledge the computing centers and personnel of the Worldwide LHC Computing Grid and other centers for delivering so effectively the computing infrastructure essential to our analyses. Finally, we acknowledge the enduring support for the construction and operation of the LHC, the CMS detector, and the supporting computing infrastructure provided by the following funding agencies: SC (Armenia), BMFWF and FWF (Austria); FNRS and FWO (Belgium); CNPq, CAPES, FAPERJ, FAPERGS, and FAPESP (Brazil); MES and BNSF (Bulgaria); CERN; CAS, MoST, and NSFC (China); MINCIENCIAS (Colombia); MSES and CSF (Croatia); RIF (Cyprus); SENESCYT (Ecuador); ERC PRG and PSG, TARISTU24-TK10 and MoER TK202 (Estonia); Academy of Finland, MEC, and HIP (Finland); CEA and CNRS/IN2P3 (France); SRNSF (Georgia); BMFTR, DFG, and HGF (Germany); GSRI (Greece); MATE and NKFIH (Hungary); DAE and DST (India); IPM (Iran); SFI (Ireland); INFN (Italy); MSIT and NRF (Republic of Korea); MES (Latvia); LMTLT (Lithuania); MOE and UM (Malaysia); BUAP, CINVESTAV, CONACYT, LNS, SEP, and UASLP-FAI (Mexico); MOS (Montenegro); MBIE (New Zealand); PAEC (Pakistan); MSHE, NSC, and NAWA (Poland); FCT (Portugal); MESTD (Serbia); MICIU/AEI and PCTI (Spain); MOSTR (Sri Lanka); Swiss Funding Agencies (Switzerland); MST (Taipei); MHESI (Thailand); TUBITAK and TENMAK (T\"{u}rkiye); NASU (Ukraine); STFC (United Kingdom); DOE and NSF (USA).
\end{acknowledgments}
\appendix
\ifthenelse{\boolean{cms@external}}{
    \section{}
}{
    \section{End Matter}
}
\label{endmatter}
\subsection{The CMS detector}
The CMS apparatus~\cite{CMS:2008xjf,CMS:2023gfb} is a multipurpose, nearly hermetic detector, designed to trigger on~\cite{CMS:2020cmk,CMS:2016ngn,CMS:2024aqx} and identify electrons, muons, photons, and charged and neutral hadrons~\cite{CMS:2020uim,CMS:2018rym,CMS:2014pgm}.
A global ``particle-flow'' algorithm~\cite{CMS:2017yfk} aims to reconstruct all individual particles in an event, combining information provided by the all-silicon tracker, and by the crystal electromagnetic and brass-scintillator hadron calorimeters, operating inside a 3.8\unit{T} superconducting solenoid, with data from the gas-ionization muon detectors interleaved with the layers of the flux-return yoke outside the solenoid.

Events of interest are selected using a two-tiered trigger system.
The first level, composed of custom hardware processors, uses information from the calorimeters and muon detectors to select events at a rate of around 100\unit{kHz} within a fixed latency of $4\unit{\mus}$~\cite{CMS:2020cmk}.
The second level, known as the high-level trigger, consists of a farm of processors running a version of the full event reconstruction software optimized for fast processing, and reduces the event rate to a few kHz before data storage~\cite{CMS:2016ngn,CMS:2024aqx}.

\subsection{Datasets and object reconstruction}

\begin{figure}[htbp]
    \centering
    \includegraphics[width=0.26\linewidth]{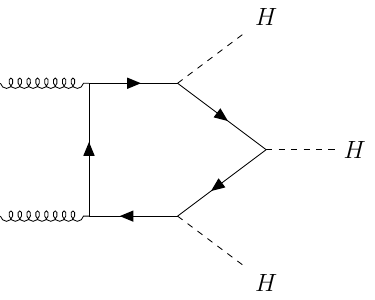}
    \includegraphics[width=0.30\linewidth]{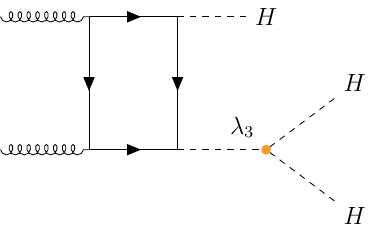}\\
    \includegraphics[width=0.30\linewidth]{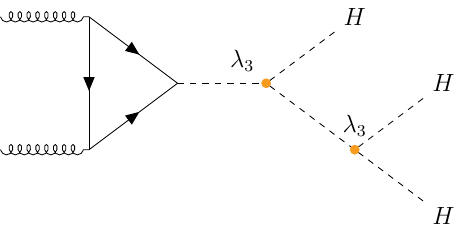}
    \includegraphics[width=0.26\linewidth]{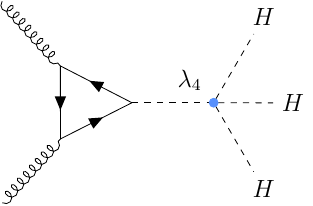}
    \caption{Representative leading-order Feynman diagrams for nonresonant \HHH production via gluon fusion.
        The last three diagrams provide direct sensitivity to the trilinear ($\lambda_3$) and quartic ($\lambda_4$) \PH self-couplings.}
    \label{fig:feynman-ggf-hhh}
\end{figure}

Signal samples for ggF \HHH production, with representative Feynman diagrams shown in Fig.~\ref{fig:feynman-ggf-hhh}, are generated at LO with \MGvATNLO~2.6.5~\cite{Alwall:2014hca}, using the model file of Ref.~\cite{Papaefstathiou:2023uum}.
Nine samples with different $(\kappa_3,\kappa_4)$ values are produced using the implementation of Ref.~\cite{Papaefstathiou:2023uum}, and intermediate points are obtained by event reweighting following Refs.~\cite{PhysRevLett.129.081802,CMS-PAS-HIG-20-011}.
The samples are normalized to the inclusive cross section at NNLO in perturbative QCD, including finite top quark mass effects~\cite{deFlorian:2019app}.
A single SM K-factor~\cite{deFlorian:2019app} is applied to all beyond-SM points, derived for the SM case to match the NNLO prediction.
The ggF \HH process is simulated at NLO using \POWHEG~2.0~\cite{POWHEG1,POWHEG2,POWHEG3,Bagnaschi:2011tu,Heinrich:2017kxx,Jones:2017giv,Buchalla:2018yce,Heinrich:2019bkc,Heinrich:2020ckp,Davies:2019dfy,Bagnaschi:2023rbx}, with additional $\kappa_3$ points constructed by reweighting and normalized to the corresponding NNLO cross sections~\cite{PhysRevD.58.115012,Shao:2013bz,PhysRevLett.111.201801,deFlorian:2015moa,PhysRevLett.117.012001,Grazzini:2018bsd,Baglio:2018lrj,PhysRevD.103.056002}.
Associated \ttHH and \VHH production processes are generated at LO with \MGvATNLO and included to study contamination effects, using the model files of Refs.~\cite{new9,new8}.

The dominant background, accounting for $\sim$95\% of events, is QCD multijet production.
It is modeled with a method fully based on control samples in data, while simulated samples produced with \MGvATNLO~2.6.5~\cite{Alwall:2014hca} at LO with up to four partons in the matrix element and binned in \HT are used exclusively for analysis optimization.
The next-to-largest background is \ttbar production, generated at NLO with \POWHEG~2.0~\cite{POWHEG1,POWHEG2,POWHEG3,Frixione:2007nw,POWHEG-TT} and normalized to the NNLO cross section using {{\textsc{Top++}}}~\cite{Czakon:2011xx,Czakon:2017wor}.
Samples of $\PW$ and $\PZ$ bosons produced in association with jets, diboson processes, and single \PH production are simulated with \MGvATNLO or \POWHEG and contribute negligibly after the final selection.

Parton showering, hadronization, and the underlying event are modeled with \PYTHIA8.205~\cite{Sjostrand:2014zea} using the CP5 tune~\cite{CMS:2019csb} and NNPDF~3.1 NNLO PDFs~\cite{Ball:2017}.
The CMS detector response is simulated with \GEANTfour~\cite{GEANT4}, and pileup interactions are overlaid and reweighted to match the distribution observed in data.

The \HHHdecay process is characterized by six \PQb quarks reconstructed as jets, balanced transverse momentum, and the absence of isolated charged leptons.
Standard jet energy corrections are applied as functions of $\eta$ and \pt~\cite{CMS:2016lmd}.
The vector \ptvecmiss is computed from the \pt sum of all reconstructed particles with PUPPI weights~\cite{CMS:2019ctu} and corrected for the scale of the reconstructed jets in the event.
Electrons and muons are reconstructed and identified according to Refs.~\cite{CMS:2020uim,CMS:2018rym,CMS:2015xaf,CMS:2016krz}, and events containing isolated leptons within the tracker and muon detector acceptance are vetoed.

Events of interest are selected using a two-tiered trigger system~\cite{CMS:2020cmk,CMS:2016ngn,CMS:2024aqx}.
In 2016, triggers required at least six jets with $\pt > 40\GeV$ and $\HT > 450\GeV$ with one $\PQb$-tagged jet, or six jets with $\pt >30\GeV$, $\HT > 400\GeV$, and two $\PQb$-tagged jets.
In 2017 (2018), triggers required at least four jets with $\pt > 40$, 45, 60, and 75\GeV, and $\HT > 300$ (330)\GeV with at least three \PQb-tagged jets.
To avoid biases, events are required to have $\HT > 450\GeV$, yielding an acceptance of $\sim$80\% for $\PH\PH\PH \to 6\PQb$ signal events.
A high-purity hadronic \ttbar CR is used to derive a residual \HT-dependent nonclosure uncertainty of 5--10\%.

In the offline selection, events are required to contain at least four small-radius jets with $\abs{\eta} < 2.4$ (2.5) and $\pt > 25\GeV$ for 2016 (2017--2018).
Small-radius jets must satisfy a tight particle-flow jet identification working point, and for $\pt < 50\GeV$ a medium pileup discriminant requirement.
Large-radius jets are selected with $\pt > 300\GeV$ and $\abs{\eta} < 2.5$.

The identification of jets originating from \PQb quarks and of \bbbar pairs is crucial for the analysis.
The \PNet algorithm~\cite{Qu:2019gqs,CMS-DP-2025-009} is employed for both small- and large-radius jets.
For small-radius jets, a 2D distribution is constructed from $p_{\prB+\prC}$, the probability for a jet to originate from a heavy-flavor quark (\PQb or \PQc), and $p_{\prB\text{vs}\prC}$, the probability for a jet to originate from a bottom rather than a charm quark.
This distribution is discretized into ten categories, following the definitions and calibrations in Ref.~\cite{CMS:2025tcl}, and all categories are used in the analysis.
For reference, the highest efficiency bin corresponds to a \PQb-tagging efficiency of $\sim$44\% for \PQb jets with $\pt > 60\GeV$, and misidentification probabilities of 0.01 and 0.10\% for light-flavor quark or gluon jets and charm jets, respectively.
The \PNet algorithm is also used for the identification of $\PH\to\bbbar$ in large-radius jets, with the discriminator output discretized into three categories corresponding to two tagging-efficiency working points, with efficiencies of 40 and 20\%, respectively, and calibrated using the $\PZ\to\bbbar$ process, as described in Ref.~\cite{CMS:2025xkn}.

\subsection{Projection for the High-Luminosity LHC program}

\begin{figure}[htbp]
    \centering
    \includegraphics[width=0.8\columnwidth]{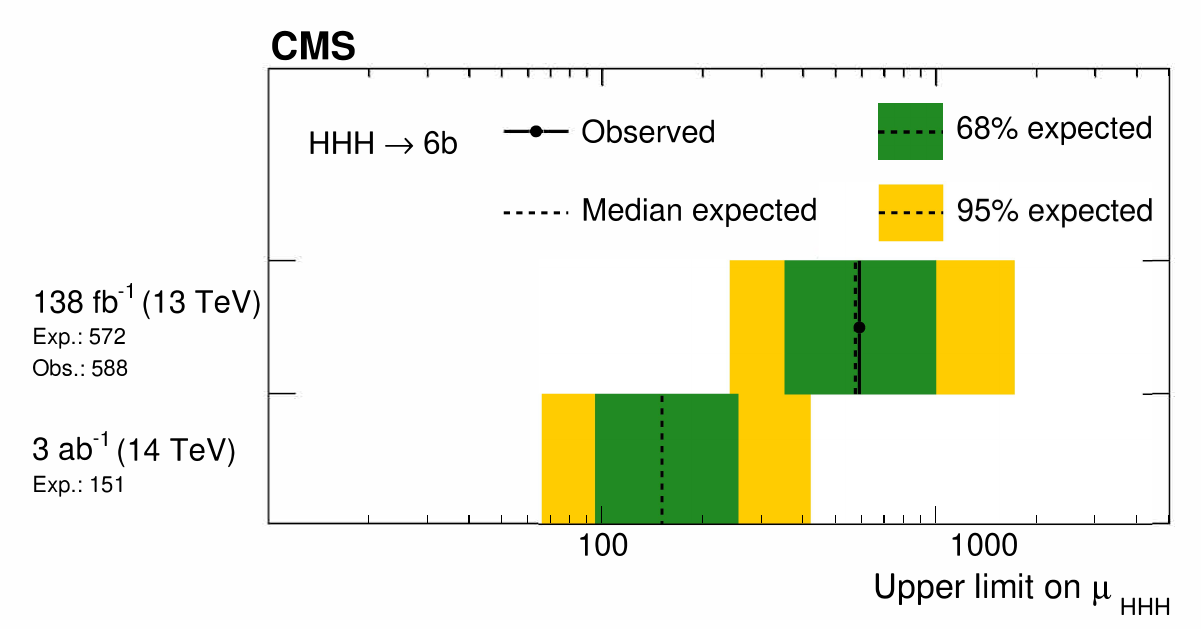}
    \caption{Observed and expected 95\% CL upper limits on the \HHH\ signal strength $\mu_{\HHH}$.
        Results are shown for the analyzed dataset of 138\fbinv of $\Pp\Pp$ collisions at $\sqrt{s}=13\TeV$, along with projections for 14\TeV and an integrated luminosity of 3\abinv.}
    \label{fig:results-summary-run2-6b}
\end{figure}

\begin{figure}[htbp]
    \centering
    \includegraphics[width=0.8\columnwidth]{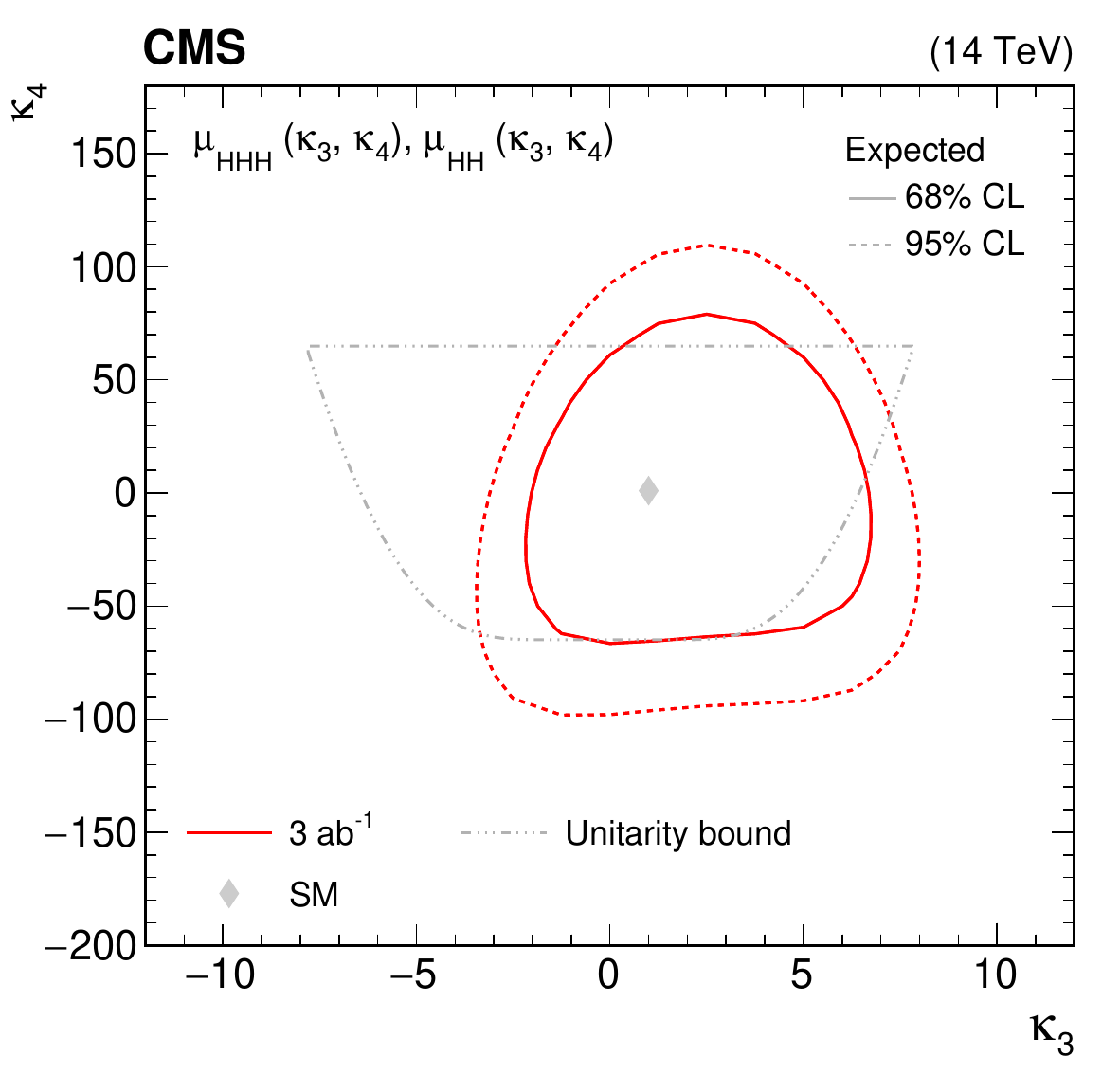}
    \caption{Expected 68 and 95\% \CL constraints in the $(\kappa_3,\kappa_4)$ plane for the $\PH\PH\PH \to 6\PQb$ analysis in the HL-LHC scenario ($\sqrt{s}=14\TeV$, 3\abinv).
        The unitarity boundary is overlaid.}
    \label{fig:hllhc-projection}
\end{figure}

Projections are derived for the HL-LHC with $\sqrt{s}=14\TeV$ and an integrated luminosity of 3\abinv under the S3 scenario~\cite{Collaboration:2928096,ATLASCMS:arXiv2504.00672}, which assumes improved \PQb-tagging performance based on \PNet optimization studies.
The projected limits include the $\kappa_4$ dependence of the \HH cross section, as done for the main result.
Figures~\ref{fig:results-summary-run2-6b} and~\ref{fig:hllhc-projection} show the current and HL-LHC sensitivities for the $\PH\PH\PH \to 6\PQb$ analysis.
Under the S3 scenario, the expected limit on the \HHH signal strength is projected to be 151 times the SM prediction, with expected constraints on the coupling modifiers $-3.2 < \kappa_3 < 7.9$ and $-92 < \kappa_4 < 99$ at 95\% \CL, assuming the other coupling modifier is 1.
Assuming only statistical uncertainties, the expected limit on the \HHH signal strength is projected to be $58$ times the SM prediction.

\ifthenelse{\boolean{cms@external}}{\clearpage}{}

\bibliography{auto_generated}
\cleardoublepage \section{The CMS Collaboration \label{app:collab}}\begin{sloppypar}\hyphenpenalty=5000\widowpenalty=500\clubpenalty=5000\cmsinstitute{Yerevan Physics Institute, Yerevan, Armenia}
{\tolerance=6000
A.~Gevorgyan\cmsorcid{0000-0003-2751-9489}, A.~Hayrapetyan, V.~Makarenko\cmsorcid{0000-0002-8406-8605}, A.~Tumasyan\cmsAuthorMark{1}\cmsorcid{0009-0000-0684-6742}
\par}
\cmsinstitute{Marietta Blau Institute for Particle Physics, Vienna, Austria}
{\tolerance=6000
W.~Adam\cmsorcid{0000-0001-9099-4341}, L.~Benato\cmsorcid{0000-0001-5135-7489}, T.~Bergauer\cmsorcid{0000-0002-5786-0293}, M.~Dragicevic\cmsorcid{0000-0003-1967-6783}, S.~Gundacker\cmsorcid{0000-0003-2087-3266}, A.K.~Guven\cmsorcid{0009-0004-5670-5138}, P.S.~Hussain\cmsorcid{0000-0002-4825-5278}, M.~Jeitler\cmsAuthorMark{2}\cmsorcid{0000-0002-5141-9560}, N.~Krammer\cmsorcid{0000-0002-0548-0985}, A.~Li\cmsorcid{0000-0002-4547-116X}, D.~Liko\cmsorcid{0000-0002-3380-473X}, M.~Matthewman, J.~Schieck\cmsAuthorMark{2}\cmsorcid{0000-0002-1058-8093}, R.~Sch\"{o}fbeck\cmsAuthorMark{2}\cmsorcid{0000-0002-2332-8784}, M.~Shooshtari\cmsorcid{0009-0004-8882-4887}, M.~Sonawane\cmsorcid{0000-0003-0510-7010}, N.~Van~Den~Bossche\cmsorcid{0000-0003-2973-4991}, W.~Waltenberger\cmsorcid{0000-0002-6215-7228}, C.E.~Wulz\cmsAuthorMark{2}\cmsorcid{0000-0001-9226-5812}
% ...
\par}
$^{1}$Also at Yerevan State University, Yerevan, Armenia\\
$^{2}$Also at Technische Universit\"{a}t Wien, Vienna, Austria\\
% ...
\end{sloppypar}
\end{document}